\documentclass[%
 aip,
 amsmath,amssymb,
 reprint,%
]{revtex4-1}
\usepackage{lineno} 
\usepackage{hyperref}
\usepackage{amsmath}  
\usepackage{amsfonts} 
\usepackage{graphicx} 
\usepackage{braket}
\usepackage{physics}
\usepackage{amsbsy}
\usepackage{textcomp}
\usepackage[gen]{eurosym}

\begin{document}



\title{Hands-On Quantum: Teaching Core Quantum Concepts With Bloch Cubes}

\author{Jeremy Levy}
\author{Chandralekha Singh}
\affiliation{Department of Physics and Astronomy, University of Pittsburgh, Pittsburgh, PA  15260 USA.}

\date{\today}

\begin{abstract}
Quantum mechanics is a notoriously abstract subject, and therefore challenging to teach at pre-college and introductory college levels.  Here we introduce the Bloch Cube, a hands-on educational tool which can illustrate key quantum concepts without equations.  A series of videos have been created showing how Bloch Cubes can be used to teach concepts such as quantum measurement, quantum dynamics, pure states versus mixed states, and quantum decoherence.  Bloch Cube states can assist in the development of more sophisticated concepts such as the Bloch Sphere, which plays a central role in the quantum mechanics of two-state systems and quantum information science.

\end{abstract}

\maketitle 

\section{Introduction} 

Quantum mechanics is an important subject that resides at the center of our physical ``theory of everything", a framework that is able to explain how energy is produced in our sun, why the periodic table is arranged as it is, why some materials are semiconductors and why they can be used to create transistors, computers, and most if not all of the technology around us~\cite{Fortier2024-wm}.  
The development of quantum science and technology, products of the ``First Quantum Revolution", played out mainly in the 20th century~\cite{MacFarlane2003-zd}.  At the epicenter of this profound revolution is a single theory and essentially one equation: the Schr\"{o}dinger equation.

Now, in the 21st century, researchers are leaning into the most mysterious and controversial predictions of quantum theory, promising a new suite of technologies--quantum computing, quantum communication, quantum sensing--that offer profound improvements over what can be achieved through conventional approaches.  At the center of the ``Second Quantum Revolution" ~\cite{MacFarlane2003-zd} is a growing concept of ``quantum information", information stored in quantum bits or ``qubits" .

\begin{figure}[hbt!]
    \centering
    \includegraphics[width=.5\linewidth]{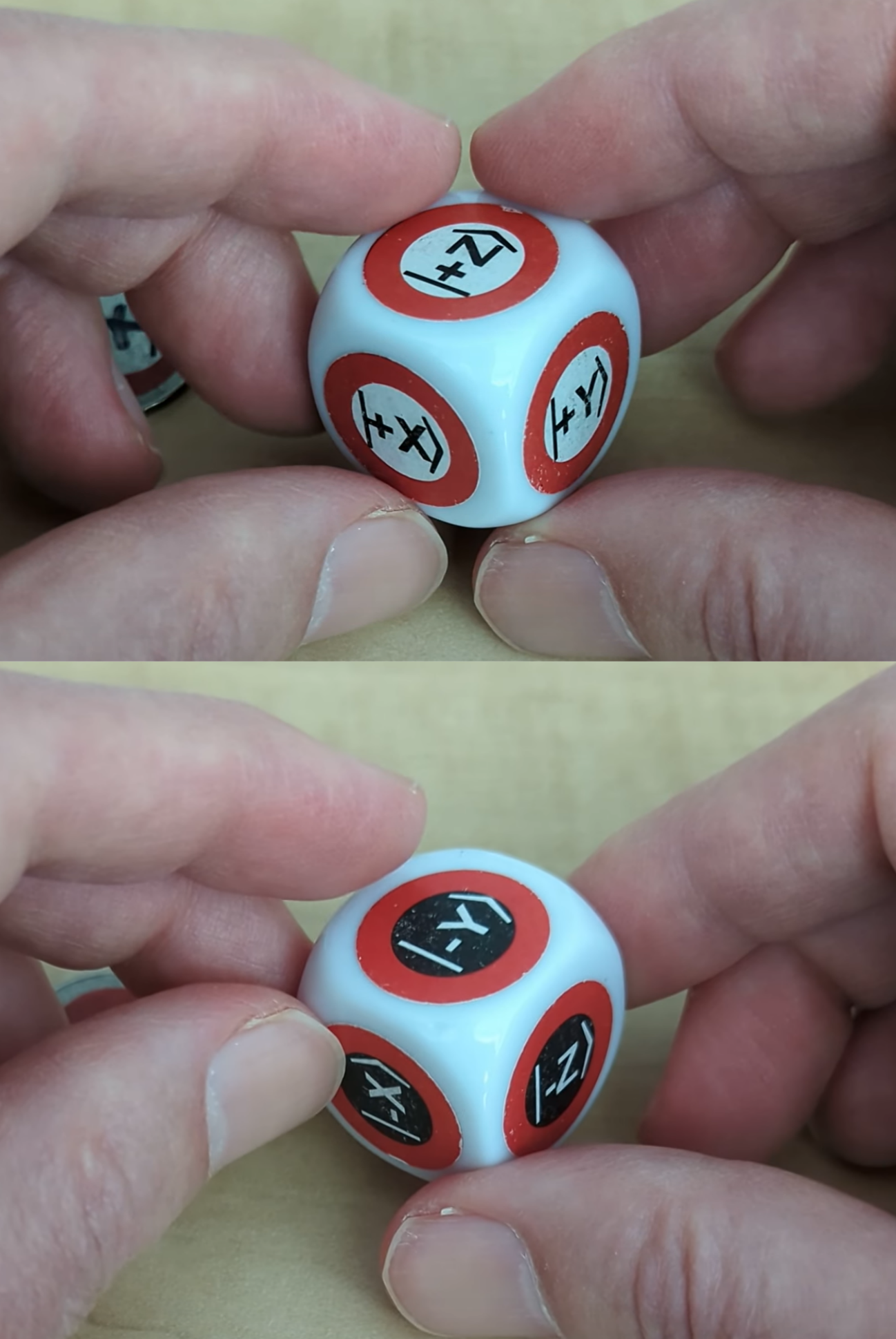}
    \caption{Front and back of a Bloch Cube.  Faces represent six quantum states of a two-dimensional quantum system.}
    \label{fig:Bloch-cube}
\end{figure}

Proper understanding and solutions of the Schr\"{o}dinger equation require sophisticated mathematical skills that are rarely mastered by students before college.
However, the importance of this subject has motivated many researchers (and ourselves included) to develop approaches that would be suitable for students who lack the technical skills thought to be necessary for understanding key quantum concepts, linear algebra, and Dirac notation~\cite{Mermin2003-ln,Grau2004-rt, Singh2022-br,Hennig2024-kx}, and topics such as quantum teleportation~\cite{Satanassi2022-gi}.
Rudolf~\cite{Rudolph2017-eq} has distilled important aspects of quantum theory, while Oss~\cite{Oss2015-me} and Dur~\cite{Dur2013-wm,Dur2014-kh,Dur2016-zx} have developed methods for teaching quantum concepts via qubits.  There are many efforts focused on teaching at high-school level, e.g., Q-12 Framework~\cite{Unknown2024-hb}. Other researchers have focused on the quantum theory of light~\cite{Bitzenbauer2020-bb,Bitzenbauer2021-xd, Stadermann2022-ya}

Quantum games have emerged as a powerful educational tool for teaching quantum mechanics concepts, with approaches ranging from board games to digital simulations. Several games focus on fundamental quantum principles like superposition and measurement, such as quantum tic-tac-toe \cite{Hoehn2014-tj, Goff2006-oz, Chiofalo2022-kt} and quantum dice implementations \cite{Yukalov2021-ul, Sassoli-de-Bianchi2013-qx}. Another category emphasizes entanglement and quantum correlations, exemplified by "Entangle Me!" \cite{Lopez-Incera2019-go} and the physically engaging "Entanglement Ball" \cite{Marckwordt2021-xk}. Digital quantum games have expanded the possibilities for interaction, with examples including Quantum Chess \cite{Cantwell2019-lw}, Quantum Minesweeper \cite{Gordon2010-ht}, and casino-style games like ``Chicago Quant'em" \cite{Gaunkar2024-nw}. Some games specifically target quantum computing concepts, such as ``Qupcakery" for learning quantum gates \cite{Liu2023-hq} and measurement simulation games \cite{Corcovilos2018-an}. Recent works have focused on quantum cryptography education \cite{Lopez-Incera2020-kj} and comprehensive reviews of quantum games in physics education \cite{Chiofalo2024-oo, Seskir2022-ds} demonstrate their growing importance as pedagogical tools.

Additionally, kinesthetic hands-on approaches are generally shown to be highly engaging for students~\cite{Kontra2015-qa, Holstermann2010-hq, Hahn2022-xr}.  
These considerations influenced the development of the Bloch Cube for teaching quantum concepts in a way that can be understood by students who are learning introductory physics. 

The authors of this article have described elsewhere~\cite{Levy2024-nq} an approach which allows high-school and first-year college students to learn core quantum concepts using notation that is modified from introductory physics courses.  An important aspect of this approach is the introduction of a hands-on quantum educational aid called the Bloch Cube, which we focus on here (FIG.~\ref{fig:Bloch-cube}).  The intended audience for this article are high school or college instructors who wish to introduce quantum concepts to students who are learning introductory physics.  The videos created here can be assigned or combined with lectures that center around the concepts covered in the videos.

The Bloch Cube is a six-sided die with specific labels $\ket{\pm z}$, $\ket{\pm x}$, $\ket{\pm y}$.  Opposite faces of the cube represent distinct quantum states, while faces that are only $90^{\circ}$ apart are related to each other via superposition. The Bloch Cube is a simplification of the concept of a Bloch sphere, restricting the allowed states from an infinite variety to just six.  This simplification still enables many key quantum concepts to be illustrated.  Crucially, students can hold a Bloch Cube and gain intuition without writing down equations.  The knowledge they gain from working with the Bloch Cube can fit into a richer framework and help students learn formal quantum mechanics at a later stage of their academic career.  The calculations enabled by the Bloch Cube are rigorous, and can be verified through formal computation with Dirac notation.

 In this article we sidestep the use of algebra to describe superposition states, and instead emphasize the concept of measurement, quantum dynamics, and other core features related to quantum mechanics of a two-state system.  A series of eight videos have been created, each of which focuses on a particular aspect of the Bloch Cube and relates its behavior to that of a two-state quantum system.  Mastery of the Bloch Cube can introduce students to key quantum concepts that are directly related to the general theory and are applicable toward an understanding of quantum information, quantum computing, and other state-of-the-art topics. 
\section{Making Your Own Bloch Cubes}
\begin{figure}[hbt!]
    \centering
    \includegraphics[width=.65\linewidth]{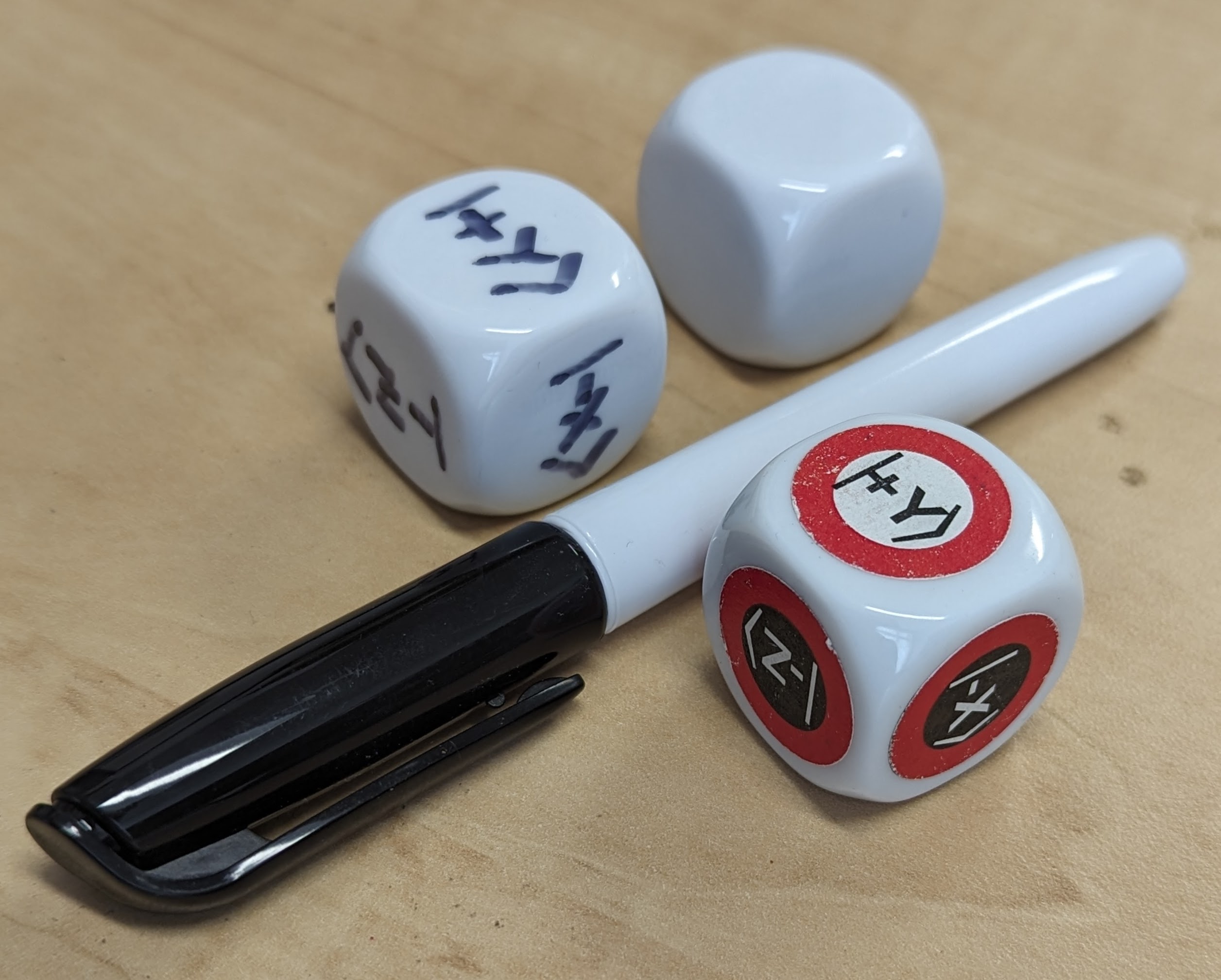}
    \caption{Make your own Bloch Cubes from blank six-sided dice.  Labels can either be printed onto stickers and applied to the faces, or they can be written with permanent marker.  The relative orientation of the labels is important for internal consistency.}
    \label{fig:Make-your-own}
\end{figure}
\begin{figure}[hbt!]
    \centering
    \includegraphics[width=0.75\linewidth]{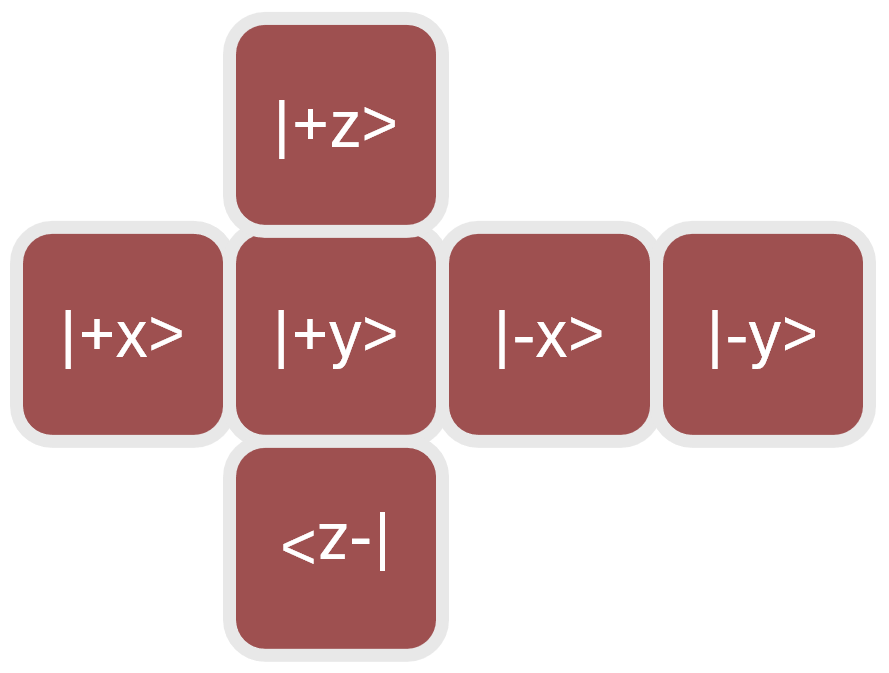}
    \caption{Expanded view of Bloch Cube faces, showing relative arrangement and orientation of the six sides.}
    \label{fig:expanded}
\end{figure}

Bloch Cubes can be fabricated using commercially available blank dice (FIG.~\ref{fig:Make-your-own}).  The six sides are marked by distinct symbols  $\ket{\pm z}$, $\ket{\pm x}$, $\ket{\pm y}$, with a conventional orientation and relationship illustrated in FIG. \ref{fig:Bloch-cube}.
To create one's own Bloch Cube, it is helpful to stick with a convention about the relative orientation of the six faces.  FIG.~\ref{fig:expanded} illustrates the convention used here and in the videos.

\section{Description of Videos}
Eight videos have been created to illustrate how Bloch Cubes can be used in a hands-on fashion to explore quantum concepts.  The videos begin by introducing the classical concept of a bit, and progress to more advanced concepts and properties of qubits.  Below we describe each of the videos. Additionally, two videos on more advanced topics cover non-equal superposition states and introduce entangled states.
The full set of videos is available
at \url{https://www.youtube.com/playlist?list=PLe-sZOiRUuMG1oQ91-IMWn-bvOYF-T96G}.

\subsubsection{A Bit about Bits}
This video illustrates the idea of a bit through the heads and tails orientation of a coin (a quarter).  The labels that describe the two states of this bit are varied in a second coin that is labeled with ``+" and ``-".

\subsubsection{Bits and Qubits}
Bits, like the head and tail of a coin, are contrasted with qubits, illustrated with the Bloch Cube.  Opposite sides of the Bloch Cube represent distinguisable states.  Unlike a coin, which has only two distinct states, the Bloch Cube has three pairs of distinct states:  $\ket{\pm z}$, $\ket{\pm x}$, $\ket{\pm y}$. States with different letters are not distinct from one another, a concept that will be made clearer in future videos. 

\subsubsection{Measuring Bits and qubits}
Measuring a bit is like uncovering a coin whose state is not known, revealing it to be heads or tails.  A similar idea exists for a qubit, except that there are (for the Bloch Cube) three types of measurements that can be performed, which can be stated in the form of questions that can be asked.  The "Z Question" asks whether the Bloch Cube face shows $\ket{+z}$ or $\ket{-z}$.  If the cube is facing $\ket{+z}$, then the answer to the Z Question will always be $+z$, and the state will be unchanged.  Similarly for $\ket{-z}$, the answer will always be $-z$, and again the state after measurement will be unchanged.  However, if the Bloch Cube is in the $\ket{\pm x}$ or $\ket{\pm y}$ state, asking the Z Question will cause the state of the Bloch Cube to change into the $\ket{+z}$ or $\ket{-z}$ state, with 50$\%$ probability.  The outcome is completely random, and this randomness is an inherent part of quantum theory.  

\subsubsection{Quantum Tomography}
The fact that the outcome of a measurement can be uncertain means that determining the quantum state cannot be performed in a single measurement.  The process of determining the initial quantum state by repeated measurements is known as quantum tomography.  This process is illustrated by trying to determine that a qubit is in the $\ket{+x}$ state.  Asking the Y Question will randomly produce $+y$ or $-y$, and asking the Z Question will randomly produce $+z$ and $-z$.  But asking the X question will always yield $+x$.  The inability to determine in a single measurement the state of an unknown qubit is related to the ``no-cloning theorem"~\cite{Unknown2024-xb}, a fundamental result in the new and expanding field of quantum information science.

\subsubsection{Quantum Dynamics}
The idea of dynamics being described by a ``flip" process is illustrated first with a coin, and then generalized to a qubit by rotating the Bloch Cube.  Rotations are considered to be applied by $\pm 90^{\circ}$ about an axis that passes through the center of one face of the cube.  The effect of these transformations can either preserve the orientation of a face, or change it.  For example, a $90^{\circ}$ clockwise rotation about the Z axis will preserve the $\ket{\pm z}$ states, but will transform the $\ket{+x}$ state into the $\ket{+y}$ state, the $\ket{+y}$ state into the $\ket{-x}$ state,  the $\ket{-x}$ state into the $\ket{-y}$ state, and the $\ket{-y}$ state into the $\ket{+x}$ state.  Similar transformations take place about the X and Y axes.  Quantum dynamics, represented by the Bloch Cube, represent solutions of the Schr\"{o}dinger equation, achieved without having to set up and solve complicated equations.

\subsubsection{Pure States and Mixed States}
The distinction between pure states and mixed states is essential for understanding real physical systems whose quantum states are not always precisely known or specified.  Introductory quantum physics courses, which generally do not cover quantum statistical mechanics, generally skip the concept of mixed states; however, this avoidance leads to confusion between ``quantum uncertainty" and ``mixture uncertainty".  Quantum uncertainty is present when the answer to the question involves a change of state, for example asking the Z Question about the $\ket{+x}$ state.  Mixture uncertainty arises when the quantum state itself is only known probabilistically.

\begin{figure}[hbt!]
    \centering
    \includegraphics[width=0.75\linewidth]{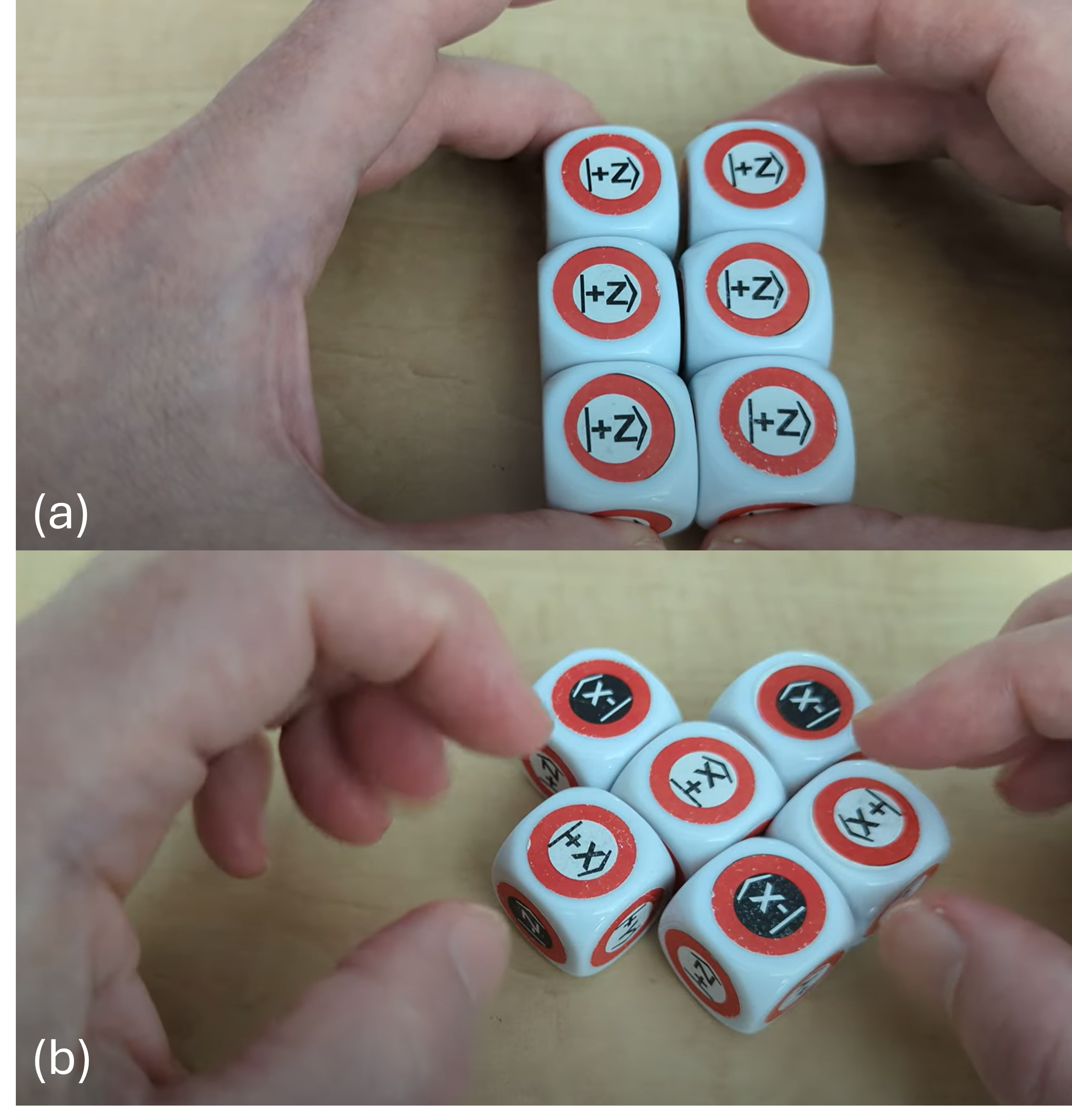}
    \caption{Example of (a) a pure state and (b) a mixed state.}
    \label{fig:pure-mixed}
\end{figure}

In the video illustration, a collection of six Bloch Cubes is used to represent a statistical ensemble of states that the quantum system can be in.  When all of the Bloch Cubes are oriented the same way (e.g., FIG.\ref{fig:pure-mixed}(a)), we have a ``pure'' state.  When there is a mixture of different states (e.g., FIG.\ref{fig:pure-mixed}(b)), we call that a ``mixed" state.  The outcome of measurements in a mixed state is calculated by considering each of the possible pure states, and weighing the answer according to their prevalence in the ensemble.  In the example given, if the Z Question were asked for the pure state (FIG.\ref{fig:pure-mixed}(a)), the answer would always be $\ket{+z}$, while asking the X Question would yield a 50-50 mixture of $\ket{+x}$ and $\ket{-x}$ states.  For the mixed state (FIG.\ref{fig:pure-mixed}(b)), half of the Bloch Cubes are in the $\ket{+x}$ state and half are in the $\ket{-x}$. 
 If one were to ask the X Question, half of the time the answer will be +1 (from the $\ket{+x}$ cubes) and half the time the answer will be -1 (from the $\ket{-x}$ cubes).  The average will be zero.  If the Z Question were asked for the mixed state, the answer would similarly average to zero.

\subsubsection{Properties of the Mixed State}
Equal mixtures of $\ket{+x}$ and $\ket{-x}$ are statistically indistinguishable from equal mixtures of $\ket{+y}$ and $\ket{-y}$.  The two types of randomness (quantum uncertainty and mixture uncertainty) are concretely illustrated.

\subsubsection{Quantum Decoherence}
Quantum decoherence is an important topic, arguably one of the most important quantum properties that distinguishes the First Quantum Revolution from the Second Quantum Revolution.  There are many ways in which coherence can be lost in a physical system, and sometimes it can be restored.  In this example, coherence is represented by a collection of four Bloch Cubes, each of which is in the $\ket{+z}$ state initially.  A rotation about the X axis switches them to the $\ket{+y}$ state.  It is then supposed that each of the Bloch Cubes will rotate, but each will spin at different rates, leading to a scrambling of the states.  This dephasing can be undone by flipping all of the cubes by $180^{\circ}$, and allowing the cubes to again spin at their different rates.  At the end, the coherence is restored, producing a quantum ``echo".  Such echoes play a central role in magnetic resonance imaging and various quantum technologies.

\section{Affordances and Limitations}
The use of a cube to represent a two-state quantum system has clear affordances.  States can be readily identified, and many of the key concepts regarding measurement uncertainty, quantum evolution, etc., can be illustrated. However, the restriction of the Bloch Cube to six states can potentially introduced unwanted difficulties.  The most obvious limitation comes from the fact that a full description of all possible quantum states would require access to the full Bloch Sphere.  Second, students may get the impression that quantum mechanics only describes two-state systems.  Other concepts, like quantum entanglement, cannot be instantiated with a Bloch Cube.  There may also be other emergent issues associated with the use of Bloch Cubes, and future educational research investigating student difficulties will be central to understanding those and figuring out how to help students learn quantum concepts early using this type of hands-on approach.

Below, we describe a specific approach that can be used to reduce student difficulties that superpositions of states can only be 50-50.  We also present an approach that is capable of describing some key aspects of quantum entanglement.  Both approaches are accompanied by videos of a style that is similar to the ones given above.

\subsection{Non-Equal Superposition States}
 It is important to realize that the restriction of states to the faces of the Bloch Cube can give students the impression that there are only six possible quantum states, with measurement outcomes that have only $0\%$, $50\%$, and $100\%$ probabilities.  This video shows by way of example how other states other than the faces of the Bloch Cube can exist, ultimately leading to a set that lives on the Bloch Sphere.
To help students interpolate between the six Bloch Cube states and the most general states of the Bloch Sphere, one can consider four pairs of ``corner states".  Opposite corners of the Bloch Cube are distinct from one another, but are not fully distinct from other pairs of corner states or from the Bloch Cube face states.  The outcome of measurements are probabilistic for all three questions X, Y, Z, and different from $0\%$, $50\%$, and $100\%$.

\subsection{Entangled States}
A single Bloch Cube does not naturally lend itself to a discussion of entangled states.  However, it is possible to describe entangled states using pairs of Bloch Cubes.  The video on entangled states shows how entanglement leads to correlations in the outcome of measurements.  The video begins by describing systems of two Bloch Cubes or qubits, showing how the outcome of measurements is independent of the order in which each qubit is measured.  The idea of an entangled state is represented by two Bloch Cubes being attached together (placed side-by-side and stuck together), and able to rotate together forward or backward.  The number of outcomes is reduced by this correlation, though there is still quantum uncertainty.  Two types of entangled states are shown, corresponding to correlated outcomes or anti-correlated outcomes.

\section{Conclusion}
We have introduced the Bloch Cube as a hands-on approach to teaching core quantum concepts.  The Bloch Cube unit described here is modular and can be incorporated in high school and college introductory courses, based upon the instructor's time constraints and preferences.  The Bloch Cube restricts the set of allowed states, which greatly simplifies understanding of quantum concepts and eliminates the need for calculus, complex numbers, and linear algebra.  Bloch Cubes can be paired up with instruction that focuses on qubits or two-state quantum systems, with emphasis on Unitary operators that map one-to-one with Bloch Cube rotations.  The cubes are themselves easy to fabricate, and can contribute to feelings of ``ownership" for students who are learning about quantum concepts for the first time.  Although the activities discussed here are primarily for high school and early college students, the Bloch Cube can also be integrated with algebraic exercises, and constitute a subset of quantum exercises that students learn in upper-level undergraduate and graduate quantum physics courses. In future, educational research will be conducted to investigate the extent to which the Bloch Cube helps students learn quantum concepts and how to reduce the difficulties they have while engaging with the Bloch Cube. Educational research would also entail development of curricular materials, e.g., worksheets, to support teachers using the Bloch Cube with their students.

\bibliography{paperpile}

\end{document}